\documentclass[12pt,preprint]{aastex}
\usepackage{graphicx}

\newcommand{\degree}{\mbox{$^{\circ}$}}
\newcommand{\am}{\mbox{\arcmin}}
\newcommand{\as}{\mbox{\arcsec}}

\def\lsim {$\rlap{\raise.4ex\hbox{$<$}}\lower.55ex\hbox{$\sim$}\,$}

\input{epsf}

\def\deg{{$^\circ$}}

\begin{document}


\title{\bf The SCUBA Legacy Catalogues:\\
Submillimetre Continuum Objects Detected by SCUBA}
\author {
James Di Francesco\altaffilmark{1,2}, 
Doug Johnstone\altaffilmark{1,2},
Helen Kirk\altaffilmark{1,2},\\
Todd MacKenzie\altaffilmark{3},
and Elizabeth Ledwosinska\altaffilmark{4}
}
\altaffiltext{1}{National Research Council of Canada, Herzberg Institute 
of Astrophysics, 5071 West Saanich Road, Victoria, BC, V9E 2E7, Canada}
\altaffiltext{2}{Department of Physics \& Astronomy, Elliot Bldg., 3800
Finnerty Road, University of Victoria, Victoria, BC, V8P 5C2, Canada}
\altaffiltext{3}{Department of Physics, University of Prince Edward Island,
550 University Avenue, Charlottetown, PE, C1A 4P3, Canada}
\altaffiltext{4}{Department of Physics, Ernest Rutherford Physics Bldg.,
McGill University, 3600 rue University, Montr\'eal, QC, H3A 2T8, Canada}


\begin{abstract}
We present the SCUBA Legacy Catalogues, two comprehensive sets of continuum 
maps (and catalogues) using data at 850 $\mu$m and 450 $\mu$m of the various 
astronomical objects obtained with the Submillimetre Common User Bolometer 
Array (SCUBA).  The Fundamental Map Dataset contains data only where superior 
atmospheric opacity calibration data were available.  The Extended Map Dataset 
is comprised of data regardless of the quality of the opacity calibration.  
Each Dataset contains 1.2\deg\ $\times$ 1.2\deg\ maps at locations where data 
existed in the JCMT archive, imaged using the matrix inversion method.  The 
Fundamental Dataset is comprised of 1423 maps at 850 $\mu$m and 1357 maps at 
450 $\mu$m.  The Extended Dataset is comprised of 1547 maps at 850 $\mu$m.  
Neither Dataset includes high sensitivity, single chop SCUBA maps of 
``cosmological fields" nor solar system objects.  Each Dataset was used to 
determine a respective Object Catalogue, consisting of objects identified 
within the respective 850 $\mu$m maps using an automated identification 
algorithm.  The Fundamental and Extended Map Object Catalogues contain 5061
and 6118 objects respectively.  Objects are named based on their respective 
J2000 position of peak 850 $\mu$m intensity.  The Catalogues provide for each 
object the respective maximum 850 $\mu$m intensity, estimates of total 850 
$\mu$m flux and size, and tentative identifications from the SIMBAD Database.  
Where possible, the Catalogues also provide for each object its maximum 450 
$\mu$m intensity and total 450 $\mu$m flux, and flux ratios.  
\end{abstract}


\section{INTRODUCTION}

In 1996, the Submillimetre Common User Bolometer Array (SCUBA) was mounted
on the 15~m diameter James Clerk Maxwell Telescope\footnote{The James Clerk
Maxwell Telescope is operated by the Joint Astronomy Centre on behalf of the
Particle Physics and Astronomy Research Council of the UK, the Netherlands
Association for Scientific Research, and the National Research Council of
Canada} (JCMT) near the summit of Mauna Kea, HI.  Since its commissioning, 
SCUBA allowed sensitive, widefield imaging of the submillimeter sky using the 
world's largest submillimetre telescope, itself located at one of the world's 
best submillimetre observing sites.  During its lifetime, SCUBA was used 
extensively by astronomers from the three JCMT partner countries (the UK, 
Canada and the Netherlands) and Hawaii, but also by many astronomers from 
other countries.  SCUBA operated for almost 9 years; in early 2005, it was 
removed from the JCMT after cryogenics and gas handling system failures.  A 
large part of the decision to remove SCUBA, rather than repair it, was that
a powerful successor instrument, SCUBA-2 (see Holland et al. 2006) will be 
installed on the JCMT in late-2007.

Throughout its productive lifetime, SCUBA was used to probe submillimeter
continuum emission from a host of various astrophysical phenomena across the 
sky observable from Mauna Kea, from objects within the Solar System to distant 
galaxies at high redshift.  SCUBA data were made available to observers of 
approved projects immediately after their acquisition and were subject to a 
proprietary period of one year after the end of the semester of observation.  
After this period, however, the data were archived at the Canadian Astronomy 
Data Centre\footnote{The Canadian Astronomical Data Centre is operated at the 
Dominion Astrophysical Observatory for the National Research Council of Canada} 
(CADC) and made available to the public.  (Students working on dissertations 
with JCMT data could have this period extended.)  SCUBA data are stored raw 
at the CADC, although preview images (made with a simple reduction) are 
available for individual files, allowing quick appraisals for data quality 
or source detection.  Over $\sim$9 years, however, many objects were observed 
over several epochs by different observers and the data spread over several 
projects and files.  Submillimeter continuum maps could be significantly 
improved by optimally combining these separate data prior to forming final 
images.

In this paper, we describe a project to image almost all SCUBA datasets, 
using raw data from all epochs, to provide an archive of images at 850 $\mu$m 
and 450 $\mu$m that were reduced consistently by a single method (i.e., the 
``matrix inversion" method described by Johnstone et al. 2000a) and using the 
most current calibrations (i.e., extinction corrections and Flux Conversion 
Factors or FCFs).  The images are themselves available for download at the 
CADC as FITS files (see \S 8 for access instructions).  We provide here, 
however, examples of some of the spectacular maps produced by SCUBA over 
its lifetime.  In addition, we present catalogues drawn from these images 
of submillimeter continuum objects mapped by SCUBA, found using an automated 
object identification program (based on the ``Clumpfind" algorithm of Williams, 
de Geus \& Blitz 1994).  No catalogue of objects at submillimeter wavelengths 
akin to the the extremely useful catalogues at near- to far-infrared 
wavelengths (e.g., the Catalog of Infrared Observations by Gezari, Schmitz 
\& Mead 1984, 1988 or the IRAS Catalogues (see Beichman et al. 1988) currently 
exists.  Since the maps produced here are derived from previous SCUBA data, 
only a relatively limited amount of sky is covered; the resulting catalogues 
are not ``all sky."  Many well-known objects and regions were mapped 
extensively with SCUBA, however.  The catalogues provide a context for 
understanding the voluminous maps that will be produced by SCUBA-2, and are 
themselves a useful planning tool for future observations with new 
submillimeter and millimeter interferometers (e.g., SMA, CARMA, ALMA).  The 
catalogues discussed here will be linked with other catalogues at the CADC.  

In the following, we describe SCUBA in \S 2, our uniform reduction procedure 
in \S 3, and a global description of the resulting maps in \S 4.   In addition,
we describe our object identification algorithm in \S 5, the SCUBA map object 
catalogues and their contents in \S 6, and several example regions in \S 7.  
A description of the available data products is provided in \S 8.  Finally, a 
summary is provided in \S 9.

\section{A BRIEF DESCRIPTION OF SCUBA}

A full description of the instrumental characteristics of SCUBA was made by 
Holland et al. (1999).  Here we describe the instrument in brief to provide
the context for the maps and catalogues.  SCUBA was built by the Royal 
Observatory Edinburgh for the James Clerk Maxwell Telescope.  It consisted 
of 128 bolometers arranged into two hexagonally packed arrays, the Long-Wave 
(LW) array with 37 bolometers and the Short-Wave (SW) array with 91 bolometers, 
as well as three additional bolometers surrounding the LW array.  Simultaneous 
illumination of the LW and SW arrays was achieved by dichroic beamsplitting, 
allowing for sampling at two wavelengths across 2.3\am\ of sky in a single 
pointing.  SCUBA's original filter set allowed for simultaneous observations 
in the LW and SW arrays either at 750 $\mu$m and 350 $\mu$m or at 850 $\mu$m 
and 450 $\mu$m.  (The 3 additional bolometers surrounding the LW array allowed 
for single-pixel observations at 1100 $\mu$m, 1300 $\mu$m and 2000 $\mu$m.)  
Most SCUBA observations, however, were made at 850 $\mu$m and 450 $\mu$m, in 
part because the filter wheel became stuck at this pair in 1997.  All SCUBA 
bolometers were cooled to $<$1 K, allowing sky background noise sensitivity 
levels to be achieved.  Subtraction of the sky was enabled by sampling 
off-target locations repeatedly during observations using the chopping 
sub-reflector of the JCMT.  Note that chopping has a profound effect on SCUBA 
data, as emission on angular scales larger than the chop throw effectively is 
spatially filtered out of the resulting maps.  At 850 $\mu$m and 450 $\mu$m, 
the resolutions of SCUBA data were represented to first order by Gaussians 
of $\sim$14\as\ and $\sim$9\as\ FWHM respectively, although significant 
``error beams" were also present, especially at 450 $\mu$m (Hogerheijde \& 
Sandell 2000; see also \S 3.1 below).  These ``error beams" must be taken 
into account when determining fluxes (see \S 4 and \S 5 below).

SCUBA was used to observe the submillimeter sky in three specific modes.  
As mentioned above, at all times the contaminating sky emission was removed
via chopping with the sub-reflector.  The first was a ``photometry" mode used 
for maximum sensitivity at the location of the central bolometer, i.e., the 
telescope effectively stared at a fixed location to maximize received signal 
from a single target.  The second was a ``jiggle" mode, used to make 
Nyquist-sampled maps across the SCUBA field-of-view, i.e., the telescope 
was moved in a fixed pattern to positions offset from each other by fractions 
of the beam to fill in spaces between the individual bolometers.  The third 
was a ``scan" mode used to make larger-scale maps at the expense of sensitivity 
at any given position, i.e., the telescope was slewed over relatively large 
distances (typically 10 arcminutes) producing strips along the sky which 
could be stitched together. By carefully choosing the angle with respect to 
the bolometer array that the telescope moved, the spacing between measurements 
Nyquist-sampled the sky.  All three modes allowed differential continuum 
intensities to be measured.  Polarized continuum emission, however, could 
also be observed across the arrays by using a rotating quartz half waveplate, 
but only in photometry or jiggle mode. 

\section{MAP DATA PROCESSING}

Since the goal of this project was to make maps and then catalogue objects 
therein, all raw jiggle and scan data from SCUBA available in the JCMT archive 
were downloaded from the CADC in May 2006.  (Photometry and polarimetry data
were ignored.)  In addition, SCUBA data taken at wavelengths other than 850 
$\mu$m or 450 $\mu$m were not retrieved.  The downloaded data consisted 
of 35455 ``SCUBA Data Files" describing for each bolometer the time of 
measurement, the location observed on the sky, and the measured voltage 
difference between that position and a specified off-position.  The 850 $\mu$m 
and 450 $\mu$m map data retrieved should have comprised all those normally 
available to the public at these wavelengths, since the instrument had ceased 
operations $>$1 year earlier.  In total size, the raw data were only 78.7 GB. 

Further culling of the raw data ensemble was necessary.  Data of objects in 
the Solar System (e.g., planets, asteroids, comets) were removed since these 
objects have time varying positions, angular sizes, and brightnesses.  Such 
files were located by visually inspecting the list of unique ``target names" 
attached to each.  (Those interested in SCUBA maps of Solar System objects at 
particular epochs can download them directly from the CADC.)  In addition, a 
small number of data attached to peculiar target names (e.g., ``whatever" or 
``reflector") were also removed from the ensemble.  Only 28534 SCUBA Data Files 
remained after culling Solar System and peculiar objects.  In size, these 
culled data were 69.9 GB.  Figure 1 shows the locations on the sky of all the 
SCUBA maps described in this paper; well-sampled areas such as the Galactic 
Plane and nearby molecular clouds like Orion and Ophiuchus are clearly visible.

Atmospheric attenuation dominates the raw voltage difference measurements
and the effect of such attenuation must be calibrated out in the data to 
obtain proper voltage levels for observed sources.  The SCUBA data were 
calibrated separately at 850 $\mu$m and 450 $\mu$m using the standard 
ORAC-DR program, part of the STARLINK package (Economou et al. 1999).  
Overall, 99.98\% of the raw files could be calibrated with ORAC-DR without 
error, and the remaining were discarded.  The baseline atmospheric opacity 
data for SCUBA calibration were obtained from a combination of skydips 
made with SCUBA itself and contemporaneous tipping scans made by a 
dedicated 225 GHz radiometer (the ``CSO 225 GHz Dipper") located at the 
nearby Caltech Submillimeter Observatory (CSO).\footnote{For the first 
half of 2003 the NRAO dipper was used as a substitute for the CSO 225 
GHz Dipper due to complications with the CSO instrument.} See Archibald 
et al. (2002) and Weferling (2005) for wide discussions of SCUBA calibration 
and how these data were tabulated for use in reducing SCUBA observations. 
Superior atmospheric correction utilizes a low-order polynomial fit in 
time to a combination of the two opacity determinations but was available 
for only 77.77\% of the SCUBA map dataset. For the rest of the data, the 
CSO 225 GHz Dipper measurement, stored in the observation header, could 
be used to estimate the sky opacity at 450 $\mu$m and 850 $\mu$m, although 
with much larger uncertainty in the conversion.

Given the importance of proper opacity correction, 850 $\mu$m and 450 
$\mu$m maps were first made with only the $\sim$78\% of data where superior 
atmospheric correction data were available.  These maps comprise the 
``Fundamental Map Dataset" and these should be referred to when interested 
in the most accurate fluxes.  To expand the scope of the maps, additional 
850 $\mu$m maps were made using {\it all} available data.  These latter data 
comprise the ``Extended Map Dataset" and these should be referred to when 
interested in the widest areal coverage.  The Extended Map Dataset does not 
include 450 $\mu$m maps because of the greater importance of accurate opacity 
calibration at shorter submillimeter wavelengths.  In the following, we treat 
the Fundamental and Extended Map Datasets equally, and provide catalogues 
derived from each.

With a nine-year lifetime, the weather conditions when SCUBA was used varied 
significantly of course.  Figure 2 shows histograms of opacity values at 850 
$\mu$m and 450 $\mu$m from data within the Fundamental Dataset, demonstrating
the spread of opacity values when SCUBA observed.

As well as calibrating the sky opacity corrections, the conversion between
voltage difference and flux must be determined. Jenness et al. (2002) showed
that over extended periods (typically semesters) during which no significant
changes to the telescope and electronics were performed, the Flux Conversion
Factor (FCF) was essentially constant with an uncertainty of approximately 
$10\%$ and $25\%$ at 850 $\mu$m and 450 $\mu$m respectively.  Most of the 
uncertainty is caused by changes in the telescope surface, due to temperature 
and gravity deformations, producing changes in the beam profile.  The 
corresponding FCF values are tabulated and available for use in data 
reduction.

To facilitate the creation of useful maps, the sky was divided into 
square-degree regions (actually each was 1.2\deg\ $\times$ 1.2\deg\ in 
extent, with 0.1\deg\ overlap with neighboring fields) using galactic 
coordinates and the individual observations comprising the Datasets 
were sorted into bins corresponding to these regions. The maps themselves, 
however, are stored in J2000 equatorial coordinates.  

Maps of each square-degree region were then made individually using the 
``mapfits" program using the respectively sorted calibrated map data as 
inputs.  Mapfits is based on the matrix inversion scheme described by 
Johnstone et al. (2000a), which produces better images from chopped data 
than techniques such as the Fourier deconvolution (e.g., Emerson et al. 
1979; Emerson 1995).  In addition, the matrix inversion method allows the 
combination of data taken with different observing setups, such as jiggle 
and scan observations.  Furthermore, data from specific bolometers are 
weighted appropriately by their respective associated noise levels.  
Finally, the image fidelity and dynamic ranges achieved by the matrix 
inversion method are good; see Johnstone et al. (2000a) for examples 
where sources of known brightness are artificially included into maps.

Previously published examples of SCUBA maps made via matrix inversion 
include those of molecular clouds in Ophiuchus (Johnstone et al. 2000b, 
Johnstone, Di Francesco \& Kirk 2004), Perseus (Kirk et al. 2006), and 
Orion (Johnstone et al. 2001; Johnstone, Matthews \& Mitchell 2006; 
Johnstone \& Bally 2006).  Additionally, L1551 in Taurus (Moriarty-Schieven 
et al. 2006) explicitly demonstrates the power of the ``mapfits" algorithm 
for bringing together heterogeneous SCUBA observations.  

Pixel sizes for the 850 $\mu$m and 450 $\mu$m maps were defined at 6\as\ 
and 3\as\ respectively.  For each square-degree region, 3 maps were created 
at each wavelength: an emission map with sky intensity pixels in Jy 
beam$^{-1}$, an error map with standard deviation values at each pixel, 
and a coverage map with each pixel containing the number of times its 
position was observed with SCUBA.  The resulting maps are projected onto 
a tangent plane associated with the center of each square-degree field.
 
A small amount of data had problems that required their respective files to 
be excised from the various square-degree regions.  These problems included: 
i) data listed as ``not a number" (NaN), ii) data that caused segmentation 
faults when running mapfits, iii) data associated with a ``wrong number of 
bolometers" and iv) data with pixels of extremely high (``infinite") noise. 
After discarding these files, the corresponding square-degree maps were 
remade using mapfits. 

Each square-degree map was further processed to remove artifacts.  First, 
noisy edges in each map were clipped.  Since such edges resulted from there
being relatively few observations at the associated pixels, the coverage 
maps were used to find pixels in the data maps at locations with less than 15 
observations, and these were clipped.  (The number maps were first smoothed 
with a Gaussian kernel of $\sigma_G$ = 7 pixels to minimize pixel-to-pixel 
variations.  Note that the square-degree maps may contain data of different
sensitivities due to differing integration times or opacity conditions during
separate observations of nearby targets.)  Second, the data maps themselves 
were smoothed with a Gaussian kernel of $\sigma_{G}$ = 1 pixel to minimize 
pixel-to-pixel noise; in effect, this increased the expected resolutions of 
the maps to $\sim$19\as\ FWHM at 850 $\mu$m and $\sim$11\as\ at 450 $\mu$m 
(but see \S 3.1 below).  Third, the data maps were flattened to remove the 
large spatial scale variations that occurred due to imperfect cancellation 
of sky signal through chopping.  Flattening was performed by first filtering 
and smoothing the original clipped data map with a Gaussian of large kernel 
size, and then subtracting this map from the smoothed and clipped data map.  
To prevent excessive bowling around bright sources, pixels with values $>5 
\times$ the median noise in the original clipped map were replaced by pixels 
with values equal to $5 \times$ the median noise.  Filtering was effective in 
reducing common artifacts where bright emission is surrounded by a bowl of 
negative pixels.  For smoothing, the filtered data map was smoothed with a 
Gaussian kernel of $\sigma_G$ = 20 pixels for the 850 $\mu$m data and 40 
pixels for the 450 $\mu$m data.  (After subtraction, edge pixels that had been 
previously clipped were reclipped.) 

To improve flux calibration, processed maps that contained three point-like 
objects, HL Tau, CRL 618 (PN G166.4-06.5) and CRL 2688 (the Egg Nebula), were 
examined.  The 850 $\mu$m and 450 $\mu$m continuum maximum intensities and 
total fluxes of these three objects were well determined by JCMT staff for 
better calibration of SCUBA data.  The comparison between the expected maximum 
intensities and those found in the processed maps at both wavelengths yielded 
correction factors that were applied respectively to all 850 $\mu$m and 450 
$\mu$m processed maps.  Table 1 lists the expected maximum intensities of all 
three objects at both wavelengths, and the maximum intensities and total fluxes 
found in the processed maps after the respective correction was made.  Small 
differences between the expected and ``observed" maximum intensities and fluxes 
still persist, but these are likely due to small variations in intrinsic source 
structure, non-centering of the object maximum intensities in a single pixel, 
and variations of observing conditions between objects.  As described below 
in \S 3, the absolute flux uncertainties of SCUBA data have been historically 
$\sim$20\% at 850 $\mu$m and $\sim$50\% at 450 $\mu$m.

Each flux-corrected and processed square-degree map was visually inspected
for quality.  In some maps containing jiggle data, periodic structures (i.e., 
ripples) were seen.  Such ripples can be introduced to maps when data obtained 
with non-standard set-ups or during times of instrumental failure are included. 
(Data were not placed into the JCMT Archive with a quality flag.)  In addition, 
such ripples may arise when jiggle data are obtained with only one chop throw 
and angle, which precludes the kind of interconnectivity between data points 
that benefits maps made by matrix inversion.  Such data are susceptible to 
amplification of the chop signal during reconstruction.  This effect typically 
does not occur over the spatial scale of a single jiggle map, but when many 
jiggle maps are combined to make a larger map, each with a single chop throw 
and angle, the opportunity increases for amplification due to degeneracy 
in reconstruction.  Unfortunately, many fields that were observed for high 
sensitivity to detect faint high-redshift galaxies, including the Hubble Deep 
Field, the Groth Strip, and the SHADES fields (the Subaru/XMM Deep Field and 
the Lockman Hole) were observed with a single chop throw and angle, and we were 
unable to produce satisfactory maps of these regions.  All square-degree maps 
entirely containing such periodic structure, including these ``cosmological" 
fields, were removed from the ensembles after visual inspection.  (Those 
interested in such fields should look at the respective papers where the data 
have been very carefully processed, e.g., see Coppin et al. 2006 for the SHADES 
fields.)  Square-degree maps containing regions of reasonable quality but 
localized regions with periodic structure (e.g., one with good scan or jiggle 
data in some locations but rippled jiggle data in other locations) were 
retained, however.  Objects found from these maps at locations of periodic 
structure were removed from catalogues after further visual inspection (see
\S 4 below).

\section{MAP RESULTS}

In the Fundamental Map Dataset, 1423 square-degree maps contain SCUBA map data 
at 850 $\mu$m and 1357 square-degree maps contain SCUBA map data at 450 $\mu$m.
(Note that 214 of the Fundamental 850 $\mu$m maps and 213 of the Fundamental 
450 $\mu$m maps contain data only in the outer 0.1\deg\ of each 1.2\deg\ 
$\times$ 1.2\deg\ field; these locations are also found within the central 
square degree in other maps of adjacent fields.)  In total, the 850 $\mu$m 
Fundamental maps contain $\sim$7.06 $\times$ 10$^{6}$ pixels for a total areal 
coverage of 19.6 square degrees.  The 450 $\mu$m Fundamental maps contain a 
total of 23.6 $\times$ 10$^{6}$ pixels for a total areal coverage of 16.4 
square degrees.  The smaller areal coverage of the 450 $\mu$m maps reflects 
the fact that at times only the 850 $\mu$m data from the telescope was stored 
during observations.  (Often this occurred during fast-scans, where the 
telescope was slewed at an accelerated rate and any 450 $\mu$m observations 
were significantly undersampled.)  In the Extended Map Dataset, 1547 
square-degree maps contain SCUBA map data at 850 $\mu$m.  (Note that 234 of 
these maps contain data only in the outer 0.1\deg\ of each 1.2\deg\ $\times$ 
1.2\deg\ field.)  These maps contain a total of 10.6 $\times$ 10$^{6}$ pixels 
for a total areal coverage of 29.3 square degrees, i.e., $\sim$50\% larger than 
in the Fundamental Map Dataset at 850 $\mu$m.  

Figures 3-6 show examples of maps assembled from the data processed in this 
effort, for low-mass star-forming regions, high-mass star-forming regions, 
nearby galaxies, and debris disks respectively.  These data, as with all 
data described here, are available for public use at the CADC. 
 
Figure 7 shows 1-D profiles of the JCMT beams at 850 $\mu$m (bottom) and 450 
$\mu$m (top), clipped to highlight the relative magnitude of the departure 
from Gaussian profiles, i.e., the error beams.  These profiles were obtained 
from slices across Fundamental Dataset data of the point-like source CRL 618
(PN G166.4-06.5).  As a common SCUBA calibrator, CRL 618 was observed numerous 
times over SCUBA's lifetime, and the data shown in Figure 7 are composites 
of all the map data of CRL 618 in the archive with proper flux calibration.  
Figures 7a and 7b show the 1-D profiles at 450 $\mu$m and 850 $\mu$m 
respectively that were obtained from maps of CRL 618 made with 1\as\ pixels.  
Figures 7c and 7d show 1-D profiles of the same object again at 450 $\mu$m 
and 850 $\mu$m respectively but obtained from maps made with 3\as\ and 6\as\ 
pixels, as in both Datasets.  In each case, the beams show clear non-Gaussian 
features but can be effectively represented by a sum of two Gaussians, a 
narrow ``primary" beam of FWHM approximately that of the expected resolution 
of the telescope at a given wavelength and smoothing and a wide ``error beam" 
of 40\as\ FWHM independent of wavelength.  For the 1\as\ maps, the 450 $\mu$m 
beam contains a primary beam of 8.5\as\ FWHM and 0.90 relative peak and the 
850 $\mu$m beam contains a primary beam of 13.5\as\ FWHM and 0.96 relative 
peak.  The values we obtain are consistent with those obtained by Hogerheijde 
\& Sandell (2000) who used data of Uranus from 1997 September, although they 
included a third, very wide, low amplitude Gaussian in their beam models at 
each wavelength.  For the 6\as\ maps, the 450 $\mu$m beam contains a primary 
beam of 11\as\ FWHM and 0.88 relative peak and the 850 $\mu$m beam contains 
a primary beam of 19.5\as\ FWHM and 0.88 relative peak.  (At both wavelengths, 
the secondary beam is of 40\as\ FHWM and 0.12 relative peak.)  These larger 
values are due to the effective smoothing that comes with using larger pixels 
but also due to the additional smoothing by $\sigma_{G}$ = 1-pixel applied to 
each map to reduce pixel-to-pixel noise.  The effective FWHMs of the beams 
in each Dataset are 17.3\as\ at 450 $\mu$m and 22.9\as\ at 850 $\mu$m.  These 
beam values are used in the computation of the observed fluxes below (see \S 
6). 

Since the maps were taken over a variety of different weather conditions and
methods, there is no common noise level representative of the entire dataset.  
Also, some maps are composites of several different observing runs, and so 
the noise level within any given map may not be uniform.  Figure 8 presents a 
histogram showing the distributions of 1~$\sigma$ rms across pixels in the 
Fundamental and Extended maps.  The 850 $\mu$m distributions have Poissonian 
characters, i.e., peaks at small values ($\sim$40 mJy beam$^{-1}$) and long 
tails to large values.  The median values of the rms at 850 $\mu$m are 71.0 
mJy beam$^{-1}$ and 76.2 mJy beam$^{-1}$ for the Fundamental and Extended maps 
respectively.   The 450 $\mu$m distribution has two peaks, however, a narrow 
one at $\sim$50 mJy beam$^{-1}$ and a broad one at $\sim$380 mJy beam$^{-1}$.  
The median value of the rms at 450 $\mu$m is 820 mJy beam$^{-1}$.  Note that 
the pixels oversample the beam at both wavelengths, so that the noise at a 
given pixel is larger than the noise within a fixed beam.  Also, for object 
identification (see \S 5 below), we use the median noise per pixel associated 
with the individual objects under investigation and not the median noise 
values of each entire square-degree map. 

Absolute flux uncertainties in the SCUBA maps were dominated by fluctuations 
of opacity above the telescope during observations and calibration.  Typical 
absolute flux uncertainties of SCUBA maps have been historically $\sim$20\% 
at 850 $\mu$m and $\sim$50\% at 450 $\mu$m (Matthews 2003), reflecting almost 
equal contributions from flux calibration and beam-shape uncertainty.  We 
adopt these uncertainties for the Catalogues in this paper.  As seen in Table 
1, the maximum intensities and fluxes of the three point-like calibrators 
HL Tau, CRL 618 (PN G166.4-06.5) and CRL 2688 (the Egg Nebula) have values 
within these uncertainties.  For further discussion of the uncertainties in 
object fluxes, see \S 7.2.

Each map was made using positional data that accompanied the respective 
SCUBA Data Files.  Pointing accuracy for SCUBA was typically $\sim$3\as\ 
and tracking accuracy was typically $\sim$1.5\as\ (H. Matthews, private 
communication).  Larger pointing offsets did occur during observations
occasionally.  For example, SCUBA data of the young stellar cluster NGC 
1333 required positional corrections of $\sim$6\as\ to line up peaks with 
data from other wavelengths (Sandell \& Knee 2001).  Given the lack of 
common positional references at other wavelengths across all map areas,
however, we have performed no positional fine-tuning on the maps.  For
further discussion of the uncertainties in positions, see \S 7.3.

Despite the care given to improving the maps here, they still may retain 
defects.  For example, some bright objects can still be surrounded by negative 
``bowls" that are obviously artificial.  In addition, map edges may still 
contain extended (positive or negative) artifacts from proper removal of sky 
emission that remain despite flattening the map. Higher accuracy determination
of fluxes and source morphologies requires significant user interaction when
map-making. The maps presented here should not be used when the highest 
precision is required, rather for such regions extreme consideration of the 
calibrations, etc., should be performed.  We expected, however, that the 
vast majority of information contained in the archival SCUBA data has been 
efficiently presented in these maps.

\section{OBJECT IDENTIFICATION}

We describe here the methods used to extract information about the objects 
detected in the 850 $\mu$m SCUBA maps.  We did not utilize the 450 $\mu$m 
maps to define objects given the lower accuracy of its flux calibration and 
the smaller number of 450 $\mu$m maps.  The identification of objects from 
submillimeter continuum emission is tricky because the emission itself can 
range in maps from being quite compact (e.g., on the order of the beam size) 
to quite extended (e.g., beyond the chop throw angular distance, although on 
these scales it becomes attenuated by the observing techniques).  In addition, 
such objects can be themselves either bright or dim and can be arranged in 
compact or diffuse associations.  

To identify objects, we applied to every 850 $\mu$m square-degree map the 
2-D ``Clumpfind" algorithm, developed first for 3-D cubes of molecular line 
data by Williams, de Geus \& Blitz (1994) and adapted for use on SCUBA 
continuum maps.  Clumpfind works by following isointensity contours within 
maps, defining objects by emission within a closed contour either 2~$\sigma$ 
below a pre-defined sensitivity limit (e.g., 5~$\sigma$) or higher if a 
neighboring object is encountered at a higher contour level.  Since objects 
are defined only in terms of closed contours, Clumpfind does not presuppose 
a particular source structure for its identifications, e.g., Gaussians.  
Since the noise level will significantly vary across any map comprised of 
observations of separate objects at different epochs, the 3 $\times$ the 
minimum noise levels of a given map were used first to define objects.  For 
each object candidate, Clumpfind returns its peak intensity and the position 
of peak intensity, as well as its total flux density and size based on the 
number of pixels within the closed contour of definition.  Clumpfind also 
produces a ``object map" that identifies pixels with specific objects.  By 
using a minimum noise threshold in every map, the algorithm was driven in a
first pass to include as many object candidates as possible; many of these 
had maximum intensities not more than a few times their local median noise 
levels, however. 

After initial identification, three criteria were applied to each object 
candidate to determine its reality as an astronomical source and improve 
the robustness of the object lists.  First, objects were discarded if their 
peak pixel values were less than 3 times the median of the noise in the pixels 
that defined it.  Second, objects were discarded if their sizes were less 
than the areal size of the effective beams (e.g., $\leq$8 pixels at 850 
$\mu$m).  Third, objects were removed if they were located near the edges 
of maps, where large scale fluctuations tended to remain even after flattening. 
For scan maps, identified objects with peaks within 25 pixels along the 
cardinal directions to the map edges were discarded if they were adjacent 
other sources that adjoined the map edge.  For jiggle maps, no such removals 
were done, given their innately smaller sizes.  Identified objects adjoining 
jiggle map edges, however, were identified as such (see \S 6), since it is 
likely they have been incompletely sampled or characterized (if indeed such 
objects are real).

Note that since Clumpfind depends on the noise characteristics of a given 
map to define objects, the objects found in the Fundamental and Extended Map 
Datasets may differ.  For example, an extended object identified as single in 
one Dataset may be identified as multiple objects in the other Dataset.  This 
dependence on noise is the reason why two object lists have been provided, 
rather than a single hybrid object list.  As an example, Figure 9 illustrates
differences between objects identified in L1688 of Ophiuchus in the Fundamental 
and Extended Datasets.

Clumpfind is by no means the perfect method for identifying objects in SCUBA 
maps.  It has, however, an attractive simplicity and generality that allowed 
it to be used on large map datasets (Johnstone et al. 2000b, etc.)  Other 
methods used to identify objects in (sub)millimeter continuum maps include 
various wavelet decomposition schemes (e.g., see Motte, Andr\'e \& Neri 1998
or Knudsen et al. 2006) or peak finding algorithms related to CLEAN-style 
deconvolution (e.g., see Enoch et al. 2006 or Young et al. 2006).  As described 
by Enoch et al., Clumpfind recovers well total flux densities in crowded 
regions of compact sources where blind aperture photometry is inappropriate.  
In addition, the Clumpfind algorithm does not unnecessarily divide up extended 
emission into multiple objects.  Clumpfind, however, likely underestimates the 
total flux densities for isolated or faint sources since significant source 
flux may reside below the pre-defined signal threshold limit.  (See \S 7.1
below for further discussions about the limitations of this technique.)

\section{THE CATALOGUES}

In this section, we describe the Catalogues based on the Fundamental and 
Extended Datasets.  Tables 2 and 3 list the Fundamental Map Object Catalogue 
(FMOC) and the Extended Map Object Catalogue (EMOC) respectively.  The FMOC 
and EMOC contain objects identified at 850 $\mu$m from the Fundamental and 
Extended Map Datasets respectively (see \S 4).  Again, the Fundamental Map 
Dataset includes only the 77.77\% of SCUBA map data where proper opacity data 
from both skydips and the CSO radiometer were available while the Extended 
Map Dataset includes all SCUBA map data deemed useable, i.e., data where only 
radiometer data were stored in the header.  (The Extended Map Dataset only 
includes maps at 850 $\mu$m, however.)  In total, the FMOC contains 5061 
objects and the EMOC contains 6118 objects, 20.4\% more than the FMOC.  

The following description of the Columns of Tables 2 and 3 applies to their 
electronic versions.  Columns 4 and 15-26 described below are not present in 
the printed versions of Tables 2 and 3.

Column 1 lists the object name, based on the position of its pixel of maximum 
brightness at 850 $\mu$m in J2000 coordinates.  The convention used is 
``JCMTS{\it n}\_JHHMMSS.S$\pm$DDMMSS" where ``JCMTS" is short for JCMT/SCUBA 
and {\it n} is either ``F" or ``E" depending on whether the object is in the 
Fundamental or Extended Catalogues respectively.  In addition, ``J" indicates
that the following coordinates are in the J2000.0 epoch.  HHMMSS.S denotes the 
hours, minutes and seconds in Right Ascension and $\pm$DDMMSS is the degrees, 
minutes and seconds in declination of the pixel of maximum intensity.  Columns 
2 and 3 list respectively the galactic coordinates\footnote{To convert from 
J2000 to galactic coordinates, the J2000.0 position of the North Galactic Pole 
was assumed to be (12$^{h}$51$^{m}$26.28$^{s}$, +27$^{\circ}$07\am 41.7\as) 
and the galactic longitude of the ascending node of the Galactic Equator was 
assumed to be 32.93192\deg, following the ICRS system values of these provided 
in the Hipparcos Catalogue (1997)} $l$ and $b$ for each object, based on the 
J2000 coordinates from Column 1.  Column 4 lists the name of the FITS file 
containing the 850 $\mu$m square-degree map from which the object was 
identified.  

Columns 5-9 list some 850 $\mu$m characteristics for each object.  Column 5
lists the object maximum 850 $\mu$m intensity in Jy beam$^{-1}$.  Column 6 
lists the object ``effective radius" in arcseconds, determined from the square 
root of the area of the object found by Clumpfind divided by $\pi$.  (Note 
that this is {\it not} the FWHM of a given object.)  Column 7 lists the median 
850 $\mu$m noise in Jy beam$^{-1}$ of all pixels associated with the object 
(defined by Clumpfind).  Column 8 lists the signal-to-noise ratio of the 
detection, i.e., the ratio of the maximum 850 $\mu$m intensity (see Column 5) 
to the median 850 $\mu$m noise (see Column 7).  Column 9 lists the object 
850 $\mu$m flux in Jy, derived as the flux of the object over its area defined 
by Clumpfind (i.e., down to a level equal to 3 $\times$ the minimum noise 
of the map of origin).  To determine 850 $\mu$m fluxes, a Gaussian beam of 
22.9\as\ FWHM was assumed (see \S 4).  

Columns 10-14 provide other 850 $\mu$m characteristics for each object.  
Column 10 lists an alternative 850 $\mu$m flux in Jy, derived as the flux 
within an alternative area, i.e., that defined by a contour of 3 $\times$ 
30 mJy beam$^{-1}$ = 90 mJy beam$^{-1}$ for all possible objects.  Determining 
fluxes for all objects within a common intensity threshold allows fluxes 
between objects to be compared more easily.  The common threshold of 30 mJy 
beam$^{-1}$ was chosen to be representative of typical noise levels of the 
850 $\mu$m maps, as seen in Figure 8.  Column 11 lists an effective radius 
of an alternative area for each object, i.e.,  where pixels had 850 $\mu$m 
intensities $\geq$ 90 mJy beam$^{-1}$.  Column 12 is a flag for the 850 
$\mu$m data.  If the maximum 850 $\mu$m intensity (Column 5) is $\geq$ 5 
$\times$ 30 mJy beam$^{-1}$ = 150 mJy beam$^{-1}$, Column 12 is blank.  If
otherwise, Column 12 lists "c" and Columns 10 and 11 list the dummy values
``-99.99" and ``-99.9" respectively.  Columns 13 and 14 list respectively 
the minimum and median noise values at 850 $\mu$m from the square-degree 
map from which the object was identified (i.e., Column 4).

Columns 15-19 list, if available, the 450 $\mu$m characteristics for each 
object.  Again, Clumpfind was not used on the 450 $\mu$m maps to define 
objects; instead 450 $\mu$m characteristics for each object are determined 
using the alternative area described above, i.e., the angular extent where 
their 850 $\mu$m intensities $\geq$ 90 mJy beam$^{-1}$.  Column 15 lists the 
median 450 $\mu$m noise in Jy beam$^{-1}$ over the alternative area of the 
object.  If the maximum 450 $\mu$m intensity within the alternative area 
(Column 11) is $>$3$\times$ the median 450 $\mu$m noise, Column 16 (a flag)
is blank, Column 17 lists the maximum 450 $\mu$m intensity in Jy beam$^{-1}$, 
Column 18 (another flag) is blank and Column 19 lists the 450 $\mu$m flux 
of the object in Jy, assuming a Gaussian beam of 17.3\as\ FWHM (see \S 4).  
Otherwise, Columns 16 and 18 list ``$<$," Column 17 lists an upper limit 
equal to 3 $\times$ the median 450 $\mu$m noise and Column 19 lists a 450 
$\mu$m flux upper limit determined by assuming each pixel within the 
alternative area (Column 11) contains a value equal to 3 $\times$ the 
median 450 $\mu$m noise.  

Columns 20-23 list, if available, the ratios of two wavelength data for each
object.  For these ratios, the 850 $\mu$m and 450 $\mu$m maps were convolved 
with beams from the other respective wavelength, to produce maps at each 
wavelength with a common beam size.  (After this convolution, both maps are 
at the same resolution and have common ``error beams.")  If an upper limit 
to the maximum 450 $\mu$m intensity is {\it not} given, Column 20 (a flag) 
is blank, Column 21 lists the ratio of maximum intensity at 450 $\mu$m to 
that at 850 $\mu$m for each object, Column 22 (a flag) is blank and Column 
23 lists the ratio of flux at 450 $\mu$m to 850 $\mu$m for each object, 
determined over the alternative area described.  If the maximum 450 $\mu$m 
intensity (Column 17) is an upper limit, Columns 20 and 22 list ``$<$" and
Column 21 lists an upper limit to the intensity ratio where the maximum 450 
$\mu$m intensity upper limit is equal to 3 $\times$ the median 450 $\mu$m 
noise, corrected to take into account the larger beam size of the convolved 
450 $\mu$m map.  Further, Column 23 lists in this case an upper limit to the 
flux ratio where the 450 $\mu$m flux upper limit is equal to that determined 
assuming each pixel in the convolved map within the alternative area contains 
a value equal to 3 $\times$ the beam-corrected median 450 $\mu$m noise.  
Note that the large uncertainties of the 850 $\mu$m and 450 $\mu$m fluxes 
make the uncertainties in their ratios accordingly large, i.e., $>$60\%. 

Column 24 provides further flags for the 450 $\mu$m data.  If the actual 
median 450 $\mu$m noise (Column 15) was $>$999 Jy beam$^{-1}$, Column 24 
lists ``n."  In addition, if no 450 $\mu$m data are present at the location 
of the object, Column 24 lists ``M."  Finally, if the maximum 850 $\mu$m 
intensity of the object is not $\geq$ 150 mJy beam$^{-1}$, Column 24 lists 
``c," as for Column 12.  In all these cases, Columns 15-19 list the dummy 
values ``-99.99."

Column 25 indicates the proximity of the object to the edge of its respective 
mapped area.  The maximum intensities and fluxes of an identified object can 
be considered accurate only if it has been sampled in its entirety across the 
sky.  To provide a sense of this accuracy, Column 19 lists either ``clear" or 
``edge" for each object.  If the former, the object was defined without any 
pixel extending to an area of the sky not mapped by SCUBA.  If the latter, the 
object extends to a map edge, and the determined fluxes should be considered
only as lower limits.

Column 26 lists potential identifications of the catalogued objects from other 
catalogues.  These were obtained from the SIMBAD astronomical database using a
bulk request for objects in the literature that were located within an 11.5\as\
radius (i.e., half the effective FWHM of the 850 $\mu$m beam) of the position 
of maximum brightness at 850 $\mu$m, as defined in Column 1.  The object chosen 
for Column 26 was that which was closest to the position of maximum 850 $\mu$m 
intensity.  Given that many astronomical objects have several names, we 
prioritized the identification of objects based on their name, or if not named, 
identification within the NGC, IC, 3C, HD, SAO, BD, or IRAS catalogues.  
(In cases of identification in several of these catalogues, the entry in Column 
26 was decided in order of how these catalogues were just listed.)  Many 
objects, however, are not found within these specific catalogues but were 
identified in various other studies.  Following the nomenclature of the SIMBAD 
database, we include in Column 26 the bibliographical abbreviation of these 
studies, along with the identification in that study.  If the SIMBAD database 
did not contain an identified object within an 11.5\as\ radius, Column 26 lists 
instead ``noMatch."  Note that extended objects can have very poorly defined 
positions (e.g., dark nebulae with positions determined from extinction maps) 
and in some cases these have been listed as ``noMatch" when its SIMBAD position 
is separated from the SCUBA 850 $\mu$m position by $>$11.5\as.  For interest, 
Figure 10 shows histograms of the numbers of objects above signal-to-noise 
thresholds of 3, 5 and 10 (see Column 8) with galactic latitude (see Column 3) 
that are listed as ``noMatch" in Column 26 in the FMOC and EMOC.  At $|b|$ $>$ 
30, 343, 189 and 75 unidentified objects are seen in the FMOC and 374, 217 and 
99 such objects are seen in the EMOC at signal-to-noise levels $\geq$ 3, 5 and 
10 respectively (see \S 7.1 for further discussion of object identification).

In the FMOC, the object with the largest maximum 850 $\mu$m intensity seen by 
SCUBA was the ``Large Molecular Heimat" associated with Sgr B2 with 242.68 Jy 
beam$^{-1}$.  Also, the object with the largest 850 $\mu$m flux seen by SCUBA 
was ``SMA 1", associated with the Orion BN/KL region at 599.6 Jy.  The total 
850 $\mu$m flux of all objects identified in the FMOC is 20868.08 Jy. 

Figure 11 shows histograms of the number of sources as a function of size 
(arcseconds), maximum 850 $\mu$m intensity (Jy beam$^{-1}$), and total 850
$\mu$um flux (Jy) for objects from the FMOC given a common sensitivity 
threshold (i.e., size determined at the 90 mJy beam$^{-1}$ level; see Columns 
10 and 11).  Figure 11a (upper left) shows that a majority of the sources have 
sizes (as measured by Clumpfind) that are resolved, with a peak in the 
distribution at $\sim$30\as.  Figure 11b (upper right) shows the maximum 
850 $\mu$m intensities.  These rise steeply toward small values, and have 
a turnover at $\sim$0.2 Jy beam$^{-1}$ likely due to the intrinsic 
sensitivities of the maps.  Figure 11c (lower left) shows the total 850 
$\mu$m fluxes with a peak near 2 Jy.  This distribution likely suffers from 
incompleteness at smaller fluxes since Clumpfind only searches out to a fixed 
intensity limit and thus underestimates the true flux of sources with low 
peak values and extents.  Figure 11d (lower right) plots the cumulative flux 
for all sources.  The integrated flux rises rapidly with lower total flux 
sources until reaching the point where the histogram turns over.  In both 
the FMOC and EMOC, the mean 850 $\mu$m flux per source is $\sim$4 Jy, while 
the median 850 $\mu$m flux is $\sim$1 Jy.

\section{ROBUSTNESS OF THE CATALOGUES}

In this section, we demonstrate the robustness of the Catalogues by comparing 
examples of Catalogue entries to various published data.  In addition, we show
by example several caveats that must be considered when interpreting data from
the SCUBA Legacy Catalogues.

\subsection{Object Identification}

Our object identification strategy identifies well locations of emission 
in each image.  For example, Johnstone al. (2002) found that most of the 67 
objects identified by Clumpfind in 850 $\mu$m maps of Orion B were largely the 
same as those identified subjectively (by eye) in the same maps by Mitchell 
et al. (2001), with differences seen only for a few very faint objects.  
Figure 12 shows the number of objects of signal-to-noise level greater than
or equal to a nominal signal-to-noise level in each Catalogue.  The number 
of objects in each Catalogue with signal-to-noise levels $\geq$3 is of course 
equal to the number of objects in each respective Catalogue, and these numbers 
drop dramatically with ever higher thresholds.  We have chosen the minimum 
local signal-to-noise level as 3 for each Catalogue since this level allows 
the inclusion of emission that appears subjectively real (by eye) in their 
parent maps.  Although this minimum signal-to-noise level is arguably low, 
recall that the objects were identified not as single pixels above this level 
but were identified from closed positive contours enclosing an area at least 
as large as the beam.  Note, however, that the Catalogues can be easily 
altered to include only objects above certain levels of signal-to-noise by 
using Column 8 of Tables 2 or 3 as a filter.

Despite its effectiveness, our object identification stratagy is not perfect.
{\bf Any judgement about a given object in the Catalogues, regardless of its 
respective signal-to-noise level, should not be made without first examining 
carefully its parent map.}  Although we have attempted to remove artifacts by 
imposing size, signal-to-noise and edge proximity criteria to all objects, 
artifacts may still remain in some maps.  The inherent heterogeneity of the 
SCUBA data means that applying uniform criteria is difficult.  Regardless of 
whatever practical criteria are applied, some artifacts will be misidentified 
as objects and some real emission will be not identified as objects. 

To illustrate our object identification strategy and demonstrate its limits, we 
provide three examples of maps from the Fundamental Map Dataset.  Figures 13-15
show 850 $\mu$m maps of L1551, M51 and NGC 7538 respectively with contours that 
delineate the boundaries of objects identified in each image by the Clumpfind 
algorithm that have passed our criteria.  Each example map shows emission that 
can be associated with actual astronomical objects.  In Figures 13 and 14, we 
see examples of low surface brightness emission that has been divided into 
multiple objects and weak objects that may be misidentified image artifacts.  
In Figures 14 and 15, we see examples of emission that remained unidentified 
as objects due to criteria imposed on each map.

In the 850 $\mu$m map of L1551 (Figure 13), 22 objects are identified.  The
three brightest are L1551 IRS 5, HL Tau and L1551-NE, located at the image 
center.  These have signal-to-noise ratios $>$70 and were easily identified 
by Clumpfind.  Another source is seen 2\am\ directly south of HL Tau with a
signal-to-noise ratio of 6.  This source corresponds to HH 30, and is seen 
because the local rms level is unusually low (i.e., 5 mJy beam$^{-1}$ ) in 
the $\sim$2\am\ diameter region around HL Tau, as this region was observed 
repeatedly as a calibrator throughout SCUBA's lifetime (see Moriarty-Schieven 
et al. 2006).  About 5\am\ northeast of L1551-NE and $\sim$2-5\am\ west of 
IRS 5, nine objects were identified with signal-to-noise ratios of 3-11, real
features that are likely associated with dust compressed by the strong outflows 
from L1551-NE or IRS 5.  Given the relatively lower surface brightness of 
these features and thus the greater influence of noise on their structure, 
Clumpfind has divided them into multiple objects.  A similar division into 
three objects is found towards the the diffuse, lower surface brightness 
feature known as L1551-MC (see Swift et al. 2005) located $\sim$7\am\ 
northwest of IRS5.  Towards the map edges, i.e., $\sim$10\am\ southwest and 
$\sim$10\am\ east of IRS5, three objects were identified respectively, all 
with low signal-to-noise ratios of 3-4.  Given the proximity of these to 
the map edge and to obvious bright artifacts at the edge (induced by imperfect 
image flattening), these objects are likely themselves artifacts.  Since it 
is difficult to know from the map alone that these latter objects are {\it 
not} artifacts, however, we retain such features as objects, leaving their 
inclusion or exclusion for further analysis up to individuals.

In the 850 $\mu$m maps of M51 (Figure 14), five objects are identified.  The 
brightest two, each with signal-to-noise ratios of $\sim$13, are located at the
galaxy nucleus in the center of the map.  The next brightest object, with a 
signal-to-noise ratio of $\sim$6, is located 5\am\ north-by-northeast of the 
M51 nucleus, and is associated with the nucleus of NGC 5195.  A fourth object, 
with a signal-to-noise ratio of $\sim$5, is associated with a bright clump to 
the southwest of the M51 nucleus along a spiral arm near the position of the 
HII region CCM 72.  Although faint emission from the spiral arms of M51 is 
clearly seen in the image, no other locations in the arms were bright enough 
relative to the local noise in this image to have been identified as objects.  
The fifth ``object" in Figure 13 with a signal-to-noise of $\sim$4, however, 
consists of a large low surface brightness feature that is likely an artifact 
of imperfect flattening in the image, similar to those seen near the edges 
of the L1551 map.  (Note the extreme high and low amplitudes seen at the map 
edges to the east and west respectively; a custom background subtraction to 
remove the edge problems and improve the detection of extended emission from 
M51 itself, was done by Meijerink et al. 2005.) 

In the 850 $\mu$m maps of NGC 7538 (Figure 15), 17 objects are identified.  
In comparison, Reid \& Wilson (2005) located 77 objects in their 850 $\mu$m
map of this region because this map had smaller pixels (2\as) and a smaller 
beam (15.3\as\ FWHM).  In addition, they constructed their map using a  
different technique (``Emerson2" reconstruction).  (Of the three examples 
discussed here, only this region had objects identified within using Clumpfind 
by other authors.)  All objects in Figure 14 are comprised of several objects 
identified by Reid \& Wilson.  The brightest three, located at the map center, 
correspond to IRS 1-3, IRS 11 and IRS 9 (respectively the SMMs 46, 48 and 
60 of Reid \& Wilson), and have signal-to-noise ratios of $>$70.  The next 
brightest object, located $\sim$2\am\ northwest of IRS 1-3 and with a 
signal-to-noise ratio of $\sim$35, is adjacent to IRS 4.  Twelve of the
remaining 13 objects have signal-to-noise ratios of $\sim$7-23 and each 
appears associated with real emission.  The last object, $\sim$2\am\ 
southeast of IRS 9 and with a signal-to-noise ratio of 3, also arguably 
appears associated with real emission; for example, Reid \& Wilson 
identified this emission with their objects SMM 69, SMM 70 and SMM 71. 
Unlike the previous two maps, no objects are identified towards the map edges 
that may be artifacts.  Notably, the map contains much weaker large-amplitude
artifacts near the edge than noted in the previous two maps.  Conversely, 
however, emission that is likely real has been not identified as an object 
given its proximity to the map edge, i.e., the emission seen $\sim$6\am\ 
west of IRS1 that is associated with SMMs 1-7 of Reid \& Wilson.  

Regarding the completeness of the Clumpfind algorithm, we stress that object 
candidates in various maps were identified down to very low noise levels in 
each map and then we used other criteria (maximum intensity vs. local median 
noise, relative location within maps) to preclude candidates from the Object 
Catalogues.  We have not attempted, however, to quantify the completeness of 
the Clumpfind algorithm, e.g., by inserting artificial sources into the maps 
to determine how well Clumpfind recovers such sources.  The objects identified 
in the Catalogues encompass a large variation of size and morphology, and the 
maps themselves can have large differences in noise both within themselves and 
between maps.  Such variety makes it difficult to make definitive tests for 
completeness across all maps.  For reference, however, we note that Enoch 
et al. (2006) performed empirical tests for completeness using Monte Carlo 
simulations of the identification of artificial Gaussian sources of various 
size in empty regions of their wide-field 1.3 mm continuum map of the Perseus 
cloud, which had a reasonably small noise level across the map ($\sim$15\%).  
Such tests defined completeness limits in mass and size, and the $\sim$100 
actual objects they identified by Clumpfind in their maps study were bounded 
on the mass-size plane by an empirically determined 10\% completeness limit, 
i.e., the level where 10\% of their artificial sources were recovered.

In summary, we believe our object identification algorithm does an effective
job of locating real emission within the 850 $\mu$m maps but it cannot be 
considered perfect.  The reality of any given object as an astronomical 
source in the Fundamental or Extended Catalogues must be considered 
carefully by those interested in these data.

\subsection{Flux Comparison}

As described earlier in \S 3, flux calibration was performed on the SCUBA 
maps using the most recent Flux Conversion Factors available in the JCMT 
archive.  In addition, we modified slightly the flux scale of all maps to 
bring the intensities of the point-source calibrators HL Tau, CRL 618 and 
CRL 2688 in line with those reported on the JCMT website (see Table 1).  In 
this section, we compare intensities obtained from our reprocessed maps with 
those from published maps that were processed by others.  In particular, we 
compare the maximum intensities of objects in the Fundamental Catalogue with 
those found in published maps after smoothing by a Gaussian of $\sigma$ = 6\as, 
binning to 6\as\ pixels, and regridding to the same pixels positions as in 
our maps using various tasks in the MIRIAD software package (Sault, Teuben 
\& Wright 1995).

Following the discussion above, we compared our Fundamental Dataset 850
$\mu$m maps of L1551 and NGC 7538 to published data of the same, provided
kindly by G. Schieven and M. Reid respectively.  From each region, a total 
of 10 or 14 objects were chosen respectively by eye from the smoothed, 
rebinned and regridded maps and the maximum intensities measured.  Figure 
16 shows the comparision between our maximum intensities and those from the 
published data at 0-3 Jy beam$^{-1}$.  For L1551 and NGC 7538, the median 
percent differences between maximum intensities of objects in the published 
data and their counterparts in the Fundamental Catalogue are 16.2\% and 
-12.8\%, i.e., within the 20\% uncertainties expected for 850 $\mu$m SCUBA 
data (see \S 3 above).  The smallest maximum intensity differences (i.e., 
$<$2\%) are found for the brightest objects in both regions (i.e., $>$3 Jy 
beam$^{-1}$; not shown in Figure 16).

Beyond different flux calibration approaches, a major component of the 
difference between the intensities in these maps is likely the difference 
between how large-scale flux variations are removed by different authors.  
As described in \S 3, the Legacy Catalogue maps have been ``flattened" by 
subtracting a very smoothed map from the original map, but different authors 
have different approaches to the problem of establishing a ``zero point" to 
SCUBA maps.  Note, however, that our technique explicitly excluded brighter
sources from smoothing prior to flattening, which may account for the smaller
differences in intensities between maps for these sources described above.

To compare the fluxes between objects, the published and Fundamental Dataset
maps were clipped according to the extents of the objects defined in the 
Fundamental Catalogues.  The percent differences in total fluxes at 850 $\mu$m 
for L1551 and NGC 7538 between the former and latter maps were 24\% and -29\% 
respectively.  Accordingly, the absolute uncertainties of fluxes at 850 $\mu$m 
of objects in the FMOC may be as large as $\pm$30\%.  Correspondingly, the 
absolute uncertainties of fluxes at 450 $\mu$m of objects in the FMOC may be 
as large as $\pm$100\%, since sky subtraction and flattening are even more 
difficult at that wavelength.

\subsection{Pointing Differences}

Pointing corrections for SCUBA were determined from short observations of 
bright, point-like calibrators.  Given variations in weather and dish surface 
conditions over the nine years SCUBA was in operation, it is thus difficult 
to assign a specific pointing uncertainty to the entire SCUBA Datasets.  
Moreover, different observers may have used different schemes over time.  
For example, note the $\sim$6\as\ offset evident in our map of NGC 1333 
relative to those made by Sandell \& Knee (2001) from early SCUBA data.

To determine a pointing uncertainty, we compare the J2000 positions of the 
three point-like calibrators, HL Tau, CRL 618 (PN G166.4-06.5) and CRL 2688 
(the Egg Nebula) as provided by SIMBAD to the J2000 position of the pixel of 
maximum intensity at 850 $\mu$m for each respective source from our Datasets.  
Not only are these particular objects compact with well defined maxima, they 
also have bright optical counterparts with fairly well established positions.  
(Note that relatively few objects detected at submillimeter wavelengths have 
optical counterparts and correspondingly have relatively poorly determined 
positions.)  Figure 17 shows the difference between the expected positions 
at (0,0) and the location of the pixel of maximum intensities.  Note that 
there is no consistent directional offset between the expected positions and 
positions of maximum intensity.  The mean magnitude of the angular offset 
between these positions is 2.7\as, or less than half the 6\as\ pixel size of 
the 850 $\mu$m maps in the Datasets, and $\sim$14\% of the 19.5\as\ FWHM 
of the narrow Gaussian component of the 850 $\mu$m beam (see Figure 7).  
(Note that 6\as\ is also equal to 1 $\sigma$ of the Gaussian representing 
the narrow component of the {\it unsmoothed} JCMT beam at 850 $\mu$m; see 
the bold dashed circle in Figure 17.)  Furthermore, the angular offsets shown 
in Figure 17 are only to the positions of the respective pixel of maximum 
intensity, which were used to identify objects in the Catalogues.  Gaussian 
fits to the 850 $\mu$m emission of each point-like source (where Gaussians 
are particularly effective) yield even closer positional coincidence.  For 
example, a mean angular offset magnitude of only 0.93\as\ is found between 
the expected positions and those of the peaks of the Gaussian fits to these 
particular objects.

\subsection{Associations with Known Objects}

Column 26 in both Catalogues (Tables 2 or 3) lists tentative associations of 
each object with those found in the SIMBAD astronomical database (i.e., known 
objects located $<$11.5\as\ from the position of maximum 850 $\mu$m intensity.)
Given the low resolution of the Dataset maps relative to optical or infrared
observations, which comprise the bulk of the SIMBAD entries, we stress that 
these associations are tentative.  For the Fundmental Catalogue, our original 
search through SIMBAD of 7031 object candidates\footnote{Note: SIMBAD source 
association was performed prior to some final culling of object candidates.
Hence the number of objects that was examined for the Fundamental Dataset was 
7031 rather than the final 5061.} resulted in potential associations with 7882 
SIMBAD objects; sometimes many SIMBAD objects were found within 11.5\as\ of
the FMOC position.  To differentiate between multiple potential associations, 
we chose objects in SIMBAD that were closest to the position of maximum 850 
$\mu$m intensity.  Of these objects, an identifier in Column 26 was chosen 
based on a either a name or entry within 8 catalogues (NGC, IC, 3C, HD, SAO, 
BD, or IRAS in decending order of priorty.)  Out of the 7031 objects, only 
1592 had associations with SIMBAD objects.

To test for false associations, all 7031 positions in the early FMOC were 
shifted north or south by 5\am\ and these new positions were run through 
SIMBAD.  These new positions resulted in only 430 or 383 potential SIMBAD
associations respectively, much less than the 7882 found earlier.  Also, out 
of the 7031 positions shifted north or south, only 356 or 322 respectively 
had associations with SIMBAD objects, again much less than the 1592 found 
earlier.  From these numbers, we surmise that the probability for false 
association in the Catalogues is relatively low, i.e., only $\sim$20\% (i.e., 
340/1600).  We note, however, that the SIMBAD database is not itself an 
all-sky survey, but rather a collection of known objects.  What is listed 
in Column 26 only indicates tentative association with previously found 
objects.

\section{DATA PRODUCTS}

The 850 $\mu$m and 450 $\mu$m square-degree maps from the Fundamental Dataset 
and the 850 $\mu$m maps from the Extended Dataset are available for download 
from the SCUBA Legacy Catalogues repository at the Canadian Astronomical 
Data Centre (CADC) at:

http://www.cadc.hia.nrc.gc.ca/community/scubalegacy. 

\noindent
Each emission map is in the standard FITS format projected onto the tangent 
plane from its center position.  Each map file is named by the galactic 
coordinates of its center position.  For example, the FITS file 
``scuba$\_$F$\_$178d6$\_$-19d8$\_$850um.emi.fits" contains the square-degree 
SCUBA 850 $\mu$m emission map from the Fundamental Dataset (F) centered at 
($l,b$) = (178.6, -19.8).  (This particular map contains a nice 850 $\mu$m 
image of L1551; see Figures 3c and 13.)  Note that files of maps from the 
Extended Dataset are identified with an ``E" instead of an ``F."

In addition to the emission maps, the repository contains other useful files, 
including the error map and coverage map corresponding to each 850 $\mu$m 
emission map.  These files are named as above but the names end in 
``850um.err.fits" or ``850um.cov.fits" respectively instead, e.g., 
``scuba$\_$F$\_$178d6$\_$-19d8$\_$850um.err.fits" or 
``scuba$\_$F$\_$178d6$\_$-19d8$\_$850um.cov.fits."
Also, the repository contains the error map corresponding to each 450 $\mu$m 
emission map.

The repository contains additional information about the objects identified in 
the 850 $\mu$m maps of the Fundamental or Extended Datasets.  For example, it
has ASCII text files containing the FMOC and EMOC, e.g., ``scuba$\_$FMOC.txt" 
and ``scuba$\_$EMOC.txt" respectively.  Finally, the repository contains object 
maps where the pixels give the numerical identifications made by Clumpfind for 
each object in respective 850 $\mu$m emission maps.  No such map was produced 
if no objects were found in the respective emission map.  These files are named 
as above but the names end in ``850um.obj.fits" instead, e.g., 
``scuba$\_$F$\_$178d6$\_$-19d8$\_$850um.obj.fits."

\section{CONCLUSIONS}

In this paper, we described the bulk processing of SCUBA map data in the JCMT
public archive, done to provide a resource of reduced 850 $\mu$m and 450 $\mu$m 
continuum data for the community and to aid future work at submillimeter and 
millimeter wavelengths.  The maps are comprised of a Fundamental Map Dataset 
at 850 $\mu$m and 450 $\mu$m and an Extended Map Dataset at only 850 $\mu$m.  
In the former Dataset, only data with superior atmospheric correction data were 
included, and in the latter, almost all available data were included.  Due to 
the specific way in which their data were collected, we do not include single
chop data from deep surveys for high-redshift galaxies, since the matrix 
inversion process did not generate satisfactory maps.  In addition, we 
described two catalogues of objects identified in the Fundamental Map and 
Extended Map Datasets, each determined using the automated ``Clumpfind" 
identification algorithm.  Maps of 850 $\mu$m and 450 $\mu$m emission as 
well as respective error, coverage, and object identification maps, 
and the catalogues, will be available for download from the CADC at 
http://www1.cadc-ccda.hia-iha.nrc-cnrc.gc.ca/jcmt/.

\acknowledgements
We thank David Bohlender and Severin Gaudet of the CADC for help with 
extracting the SCUBA map data from the JCMT archive.  We thank Wayne 
Holland, Walter Gear and Ian Robson for their assistance in placing
SCUBA on the JCMT, and we thank G\"oran Sandell, Gerald Schieven, and 
Henry E. Matthews for their assistance with SCUBA over its lifetime.
In addition, we thank Gary Davis, Director JCMT, Antonio Chrysostomou,
Deputy Director JCMT, the staff at the JAC and ROE, PPARC, many JCMT 
observers and Telescope Support Specialists over the years for their
support.  Finally, we thank Russell Redman, John Ouellette, Pat Dowler 
and Sharon Goliath for assistance in ingesting the data into CADC for
wide distribution.  This research has made extensive use of the SIMBAD 
database, operated at CDS, Strasbourg, France.

\clearpage

\begin{deluxetable}{ccccccc}
\tablecolumns{7}
\footnotesize
\tablecaption{\bf Expected vs. Observed Fluxes of Point-like Calibrators
\label{tab1}}
\tablewidth{0pt}
\tablehead{
\colhead{} & 
\colhead{Expected} & 
\colhead{Observed} &
\colhead{Observed} &
\colhead{Expected} & 
\colhead{Observed} &
\colhead{Observed} \\
\colhead{} & 
\colhead{850 $\mu$m} & 
\colhead{850 $\mu$m} & 
\colhead{850 $\mu$m} & 
\colhead{450 $\mu$m} & 
\colhead{450 $\mu$m} & 
\colhead{450 $\mu$m} \\
\colhead{Name} & 
\colhead{Peak} & 
\colhead{Peak} &
\colhead{Flux} &
\colhead{Peak} & 
\colhead{Peak} &
\colhead{Flux} \\
\colhead{} & 
\colhead{(Jy beam$^{-1}$)} & 
\colhead{(Jy beam$^{-1}$)} & 
\colhead{(Jy)} & 
\colhead{(Jy beam$^{-1}$)} & 
\colhead{(Jy beam$^{-1}$)} & 
\colhead{(Jy)}
}
\startdata
HL Tau     & 2.35 $\pm$ 0.08 & 2.4 & 2.1\tablenotemark{a} / 1.9\tablenotemark{b}  & 9.4 $\pm$ 1.3  & 12. & 8.6 \\
CRL 618    & 4.6 $\pm$ 0.2   & 4.4 & 5.0\tablenotemark{a} / 4.1\tablenotemark{b} & 10.9 $\pm$ 0.9 & 8.2 & 8.9 \\
CRL 2688   & 5.9 $\pm$ 0.2   & 5.8 & 5.2\tablenotemark{a} / 5.1\tablenotemark{b} & 22.0 $\pm$ 2.7 & 24. & 19. \\
\enddata
\tablenotetext{a}{Flux calculated from full area of object.}
\tablenotetext{b}{Flux calculated from area within 90 mJy beam$^{-1}$ contour.}
\end{deluxetable}

\clearpage

\begin{deluxetable}{ccccccccccccc} 
\rotate
\tablecolumns{13}
\tablewidth{0pt}
\tabletypesize{\scriptsize}
\setlength{\tabcolsep}{0.02in}
\tablecaption{Fundamental Map Object Catalogue} 
\tablehead{ 
 & & & \colhead{850 $\mu$m} & & \colhead{850 $\mu$m} & & & \colhead{Alternative} & \colhead{Alternative} & & \colhead{850 $\mu$m} & \colhead{850 $\mu$m} \\
\colhead{Object} & \colhead{Galactic} & \colhead{Galactic} & \colhead{Maximum} & \colhead{Object} & \colhead{Median} & \colhead{Peak} & \colhead{850 $\mu$m} & \colhead{850 $\mu$m} & \colhead{Object} & \colhead{850 $\mu$m}& \colhead{Minimum} & \colhead{Median} \\
\colhead{Identifier}  &  \colhead{Longitude}  &  \colhead{Latitude} &  \colhead{Intensity}  &  \colhead{Size}  &  \colhead{RMS}  &  \colhead{S/N}  &  \colhead{Flux}  &  \colhead{Flux}  &  \colhead{Size}  & Flag & \colhead{Map RMS}  & \colhead{Map RMS} \\
& \colhead{(\degree)} & \colhead{(\degree)} & \colhead{(Jy beam$^{-1}$)} & \colhead{(\as)} & \colhead{(Jy beam$^{-1}$)} & & \colhead{(Jy)} & \colhead{(Jy beam$^{-1}$)} & \colhead{(\as)} & & \colhead{(Jy beam$^{-1}$)} & \colhead{(Jy beam$^{-1}$)}}
\startdata 
JCMTSF\_J000134.7$+$231250 & 108.3034 & -38.2374 & 0.06 &  18.2 &  0.02 &  3.0   & 0.08  & -99.99 & -99.9 &c&  0.006 & 0.019  \\
JCMTSF\_J000136.0$+$231308 & 108.3111 & -38.2338 & 0.08 &  16.6 &  0.02 &  4.0   & 0.07  & -99.99 & -99.9 &c&  0.006 & 0.019  \\
JCMTSF\_J000136.8$+$231056 & 108.3035 & -38.2701 & 0.27 &  10.7 &  0.02 &  16.2  &  0.06 &   0.05 &   6.8 & &  0.006 & 0.019  \\
JCMTSF\_J000137.3$+$231332 & 108.3193 & -38.2286 & 0.08 &  14.3 &  0.02 &  4.0   & 0.05  & -99.99 & -99.9 &c&  0.006 & 0.019  \\
JCMTSF\_J000137.7$+$231232 & 108.3160 & -38.2451 & 0.06 &  30.7 &  0.02 &  3.8   & 0.17  & -99.99 & -99.9 &c&  0.006 & 0.019 \\
JCMTSF\_J000138.6$+$231050 & 108.3115 & -38.2734 & 0.11 &   8.3 &  0.02 &  6.2   & 0.02  &   0.01 &   3.4 & &  0.006 & 0.019  \\
JCMTSF\_J000138.9$+$233102 & 108.4172 & -37.9472 & 0.55 &  12.7 &  0.07 &  7.9   & 0.18  &   0.16 &   9.6 & &  0.006 & 0.019  \\
JCMTSF\_J000141.5$+$232944 & 108.4227 & -37.9706 & 0.04 &  12.2 &  0.01 &  5.0   & 0.03  & -99.99 & -99.9 &c&  0.006 & 0.019  \\
JCMTSF\_J000141.6$+$231044 & 108.3251 & -38.2778 & 0.23 &  10.7 &  0.02 &  10.0  &  0.04 &   0.02 &   4.8 & &  0.006 & 0.019  \\
JCMTSF\_J000142.9$+$231138 & 108.3360 & -38.2645 & 0.06 &  28.9 &  0.02 &  3.3   & 0.14  & -99.99 & -99.9 &c&  0.006 & 0.019  \\
JCMTSF\_J000146.8$+$232914 & 108.4451 & -37.9836 & 0.73 &  10.1 &  0.01 &  70.0  &  0.14 &   0.12 &   7.6 & &  0.006 & 0.019  \\
JCMTSF\_J000358.6$+$683507 & 118.6019 &   6.1135 & 0.86 &  33.0 &  0.07 &  11.7  &  1.91 &   1.91 &  33.0 & &  0.048 & 0.076  \\
JCMTSF\_J000522.7$+$671751 & 118.4975 &   4.8232 & 0.36 &  39.8 &  0.08 &  4.4   & 1.24  &   1.05 &  33.5 & &  0.023 & 0.040  \\
JCMTSF\_J000754.5$+$352220 & 113.0128 & -26.6597 & 0.06 &   8.3 &  0.01 &  4.3   & 0.02  & -99.99 & -99.9 &c&  0.008 & 0.021  \\
JCMTSF\_J000953.3$+$255525 & 111.3672 & -36.0133 & 0.06 &  12.2 &  0.01 &  5.0   & 0.04  & -99.99 & -99.9 &c&  0.011 & 0.019  \\
JCMTSF\_J000955.1$+$255356 & 111.3693 & -36.0389 & 0.69 &  11.2 &  0.09 &  7.3   & 0.17  &  0.15  & 8.3   & &  0.011 & 0.019  \\
\enddata 
\tablecomments{Table 2 is published in its entirety in the electronic edition 
of the {\it Astrophysical Journal}.  A portion is shown here for guidance 
regarding its form and content.  The online only version contains more columns,
including the names of the source map of each entry and the 450 $\mu$m data.}
\end{deluxetable} 

\clearpage

\begin{deluxetable}{ccccccccccccc} 
\rotate
\tablecolumns{13}
\tablewidth{0pt}
\tabletypesize{\scriptsize}
\setlength{\tabcolsep}{0.02in}
\tablecaption{Extended Map Object Catalogue} 
\tablehead{ 
 & & & \colhead{850 $\mu$m} & & \colhead{850 $\mu$m} & & & \colhead{Alternative} & \colhead{Alternative} & & \colhead{850 $\mu$m} & \colhead{850 $\mu$m} \\
\colhead{Object} & \colhead{Galactic} & \colhead{Galactic} & \colhead{Maximum} & \colhead{Object} & \colhead{Median} & \colhead{Peak} & \colhead{850 $\mu$m} & \colhead{850 $\mu$m} & \colhead{Object} & \colhead{850 $\mu$m}& \colhead{Minimum} & \colhead{Median} \\
\colhead{Identifier}  &  \colhead{Longitude}  &  \colhead{Latitude} &  \colhead{Intensity}  &  \colhead{Size}  &  \colhead{RMS}  &  \colhead{S/N}  &  \colhead{Flux}  &  \colhead{Flux}  &  \colhead{Size}  & Flag & \colhead{Map RMS}  & \colhead{Map RMS} \\
& \colhead{(\degree)} & \colhead{(\degree)} & \colhead{(Jy beam$^{-1}$)} & \colhead{(\as)} & \colhead{(Jy beam$^{-1}$)} & & \colhead{(Jy)} & \colhead{(Jy beam$^{-1}$)} & \colhead{(\as)} & & \colhead{(Jy beam$^{-1}$)} & \colhead{(Jy beam$^{-1}$)}}
\startdata 
JCMTSE\_J000134.7$+$231250 & 108.3034 & -38.2374 & 0.06 &  18.2 &  0.02  &   3.0  &  0.08 &  -99.99 & -99.9 & c &  0.006 & 0.019 \\ 
JCMTSE\_J000136.0$+$231308 & 108.3111 & -38.2338 & 0.06 &  16.6 &  0.02  &   3.0  &  0.07 &  -99.99 & -99.9 & c &  0.006 & 0.019 \\
JCMTSE\_J000136.8$+$231056 & 108.3035 & -38.2701 & 0.27 &  10.7 &  0.02  &  16.2  &  0.06 &    0.05 &   6.8 &   &  0.006 & 0.019 \\
JCMTSE\_J000137.3$+$231332 & 108.3193 & -38.2286 & 0.08 &  14.3 &  0.02  &   4.0  &  0.05 &  -99.99 & -99.9 & c &  0.006 & 0.019 \\
JCMTSE\_J000137.7$+$231232 & 108.3160 & -38.2451 & 0.06 &  30.7 &  0.02  &   3.8  &  0.17 &  -99.99 & -99.9 & c &  0.006 & 0.019 \\
JCMTSE\_J000138.6$+$231050 & 108.3115 & -38.2734 & 0.11 &   8.3 &  0.02  &   6.2  &  0.02 &    0.01 &   3.4 &   &  0.006 & 0.019 \\
JCMTSE\_J000138.9$+$233102 & 108.4172 & -37.9472 & 0.55 &  12.7 &  0.07  &   7.9  &  0.18 &    0.16 &   9.6 &   &  0.006 & 0.019 \\
JCMTSE\_J000141.5$+$232944 & 108.4227 & -37.9706 & 0.04 &  12.2 &  0.01  &   5.0  &  0.03 &  -99.99 & -99.9 & c &  0.006 & 0.019 \\
JCMTSE\_J000141.6$+$231044 & 108.3251 & -38.2778 & 0.23 &  10.7 &  0.02  &  10.0  &  0.04 &    0.02 &   4.8 &   &  0.006 & 0.019 \\
JCMTSE\_J000142.9$+$231138 & 108.3360 & -38.2645 & 0.06 &  28.3 &  0.02  &   3.3  &  0.13 &  -99.99 & -99.9 & c &  0.006 & 0.019 \\
JCMTSE\_J000146.8$+$232914 & 108.4451 & -37.9836 & 0.73 &  10.7 &  0.01  &  58.3  &  0.16 &    0.14 &   8.3 &   &  0.006 & 0.019 \\
JCMTSE\_J000358.6$+$683507 & 118.6019 &   6.1135 & 0.82 &  42.1 &  0.06  &  14.4  &  3.16 &    3.16 &  42.1 &   &  0.055 & 0.078 \\
JCMTSE\_J000401.6$+$683901 & 118.6184 &   6.1765 & 0.27 &  30.8 &  0.06  &   4.5  &  1.03 &    1.03 &  30.8 &   &  0.055 & 0.078 \\
JCMTSE\_J000402.9$+$683619 & 118.6120 &   6.1319 & 0.48 &  46.4 &  0.06  &   7.9  &  3.14 &    3.14 &  46.4 &   &  0.055 & 0.078 \\
JCMTSE\_J000411.6$+$683837 & 118.6322 &   6.1672 & 0.46 &  41.6 &  0.06  &   7.6  &  2.24 &    2.24 &  41.6 &   &  0.055 & 0.078 \\
JCMTSE\_J000413.9$+$683619 & 118.6286 &   6.1289 & 0.44 &  52.2 &  0.06  &   7.2  &  4.34 &    4.34 &  52.2 &   &  0.055 & 0.078 \\
\enddata 
\tablecomments{Table 3 is published in its entirety in the electronic edition 
of the {\it Astrophysical Journal}.  A portion is shown here for guidance 
regarding its form and content.  The online only version contains more columns,
including the names of the source map of each entry and the 450 $\mu$m data.}
\end{deluxetable} 

\clearpage

\begin{figure}
\figurenum{1}
\epsscale{1.0}
\vspace*{-4.0cm}
\caption{Distribution of SCUBA mapping locations across the sky, excluding
observations of Solar System objects (see text), where each J2000 position
observed is denoted by a cross.  Some crosses are darker than others because
they are the superposition of several observations made in close proximity.}
\end{figure}

\clearpage

\begin{figure}
\figurenum{2}
\epsscale{1.0}
\vspace*{-2.0cm}
\vskip -4.0cm
\caption{Histograms of optical depth (``tau") values measured at 850 $\mu$m 
(upper panel) and 450 $\mu$m (lower panel) for data within the Fundamental 
Dataset.  Note that the histograms show only the majority of measured values
and the bins at the extreme right contain the totals at values greater than 
or equal to the respective extremes.}
\end{figure}

\clearpage

\begin{figure}
\figurenum{3}
\epsscale{1.0}
\vspace*{-2.0cm}
\caption{Examples of SCUBA 850 $\mu$m observations of low-mass star-forming
regions from the Extended Dataset.  Greyscale ranges are chosen to bring out 
low level features in the maps.  a) The L1688 cluster region in the central 
part of the Ophiuchus molecular cloud with greyscale ranging from -0.2 Jy 
beam$^{-1}$ to 1.1 Jy beam$^{-1}$.  b) A central part of the Taurus molecular 
cloud with greyscale ranging from -0.1 Jy beam$^{-1}$ to 0.6 Jy beam$^{-1}$.
c) The L1551 cloud, south of the Taurus molecular cloud with greyscale ranging
from -0.1 Jy beam$^{-1}$ to 0.6 Jy beam$^{-1}$.  d) A region of young star 
formation south of IC 348 in the Perseus molecular cloud with greyscale ranging 
from -0.1 Jy beam$^{-1}$ to 0.6 Jy beam$^{-1}$.  e) The isolated starless 
core Barnard 68 with greyscale ranging from -0.04 Jy beam$^{-1}$ to 0.21 Jy 
beam$^{-1}$.  f) The bright, clustered protostellar sources of the Serpens 
molecular cloud with greyscale ranging from -0.6 Jy beam$^{-1}$ to 3.2 Jy 
beam$^{-1}$}.
\end{figure}

\clearpage

\begin{figure}
\figurenum{4}
\epsscale{1.0}
\vspace*{-1.0cm}
\vskip -1.0cm
\caption{Examples of SCUBA 850 $\mu$m observations of high-mass star-forming
regions from the Extended Dataset.  Greyscale ranges are chosen to bring out
low level features in the maps.  a) Inner part of the W3 molecular cloud with 
greyscale ranging from -0.2 Jy beam$^{-1}$ to 1.1 Jy beam$^{-1}$.  b) The NGC 
2068 region of the Orion B molecular cloud (including the Horsehead Nebula) 
with greyscale ranging from -0.2 Jy beam$^{-1}$ to 1.1 Jy beam$^{-1}$.  c) 
The NGC 6334 filament with greyscale ranging from -2.0 Jy beam$^{-1}$ to 11 
Jy beam$^{-1}$.  d) The DR 21 region with greyscale ranging from -0.4 Jy 
beam$^{-1}$ to 2.1 Jy beam$^{-1}$}.
\end{figure}

\clearpage

\begin{figure}
\figurenum{5}
\epsscale{1.0}
\vspace*{-1.0cm}
\caption{Examples of SCUBA 850 $\mu$m observations of nearby galaxies from
the Extended Dataset.  In all panels, the greyscale ranges from -0.1 Jy 
beam$^{-1}$ to 0.32 Jy beam$^{-1}$. a) The Whirlpool Galaxy (M51).  b) The 
central part of the colliding Antennae galaxies (NGC 4038/4039). c) The 
nearby galaxy NGC 1068.  d) The peculiar galaxy Arp 220.}
\end{figure}

\clearpage

\begin{figure}
\figurenum{6}
\epsscale{1.0}
\vspace*{-3.0cm}
\caption{Examples of SCUBA 850 $\mu$m observations of debris disks around main
sequence stars.  In all panels, the greyscale ranges from -0.006 to 0.04 Jy
beam$^{-1}$. a) $\alpha$ PsA (Fomalhaut).  b) $\beta$ Pic.  c) AU Mic.  d)
$\alpha$ Lyr (Vega).  e) $\epsilon$ Eri.  f) $\eta$ Crv.}
\end{figure}

\clearpage

\begin{figure}
\figurenum{7}
\epsscale{1.0}
\vspace*{-4.0cm}
\caption{The low-level structure in the JCMT beams at 450 $\mu$m (upper two 
panels) and at 850 $\mu$m (lower two panels), from numerous observations of 
the calibrator source CRL 618 (PN G166.4-06.5).  In each plot, the dots show 
pixel values normalized to the peak intensity.  Solid lines show Gaussian 
profiles fit to the profile by eye from the summation of the two other Gaussian 
profiles shown as dashed lines.  Panels on the left are from images of CRL 618 
made with 1\as\ pixels while panels on the right are taken from Fundamental 
Dataset images (i.e., 3\as\ pixels at 450 $\mu$m and 6\as\ pixels at 850 
$\mu$m.)} 
\end{figure}

\clearpage

\begin{figure}
\figurenum{8}
\epsscale{1.0}
\vspace*{-0.1cm}
\caption{Histograms showing the distribution of 1 $\sigma$ rms per pixel from
all pixels in the 850 $\mu$m Extended maps (dotted bold line), 850 $\mu$m 
Fundamental maps (solid bold line) and the 450 $\mu$m Fundamental maps (solid 
normal line).  The distributions have been truncated at 1.0 Jy beam$^{-1}$.}
\end{figure}

\clearpage

\begin{figure}
\figurenum{9}
\epsscale{1.0}
\vspace*{-0.1cm}
\vskip -2.0cm
\caption{Differences in L1688 objects in the Fundamental (top) and Extended 
(bottom) Datasets where different objects in each Dataset are shown by 
different shades of grey.} 
\end{figure}

\clearpage

\begin{figure}
\figurenum{10}
\epsscale{1.0}
\vspace*{-0.1cm}
\caption{Logarithmic histograms of the numbers of unidentified objects with 
galactic latitude in the FMOC (upper panel) and the EMOC (lower panel).  Such 
objects are those labelled ``noMatch" in Column 26 of Tables 2 and 3.}
\end{figure}

\clearpage

\begin{figure}
\figurenum{11}
\epsscale{1.0}
\vspace*{-0.1cm}
\caption{Summary plots of the FMOC.  The top left panel is a histogram of 
the logarithmic distribution of the alternative effective sizes in arcseconds 
of all objects down to the 90 mJy beam$^{-1}$ contour (Column 11 of Table 2).  
The top right panel is a histogram of the logarithmic distribution of the 
alternative 850 $\mu$m fluxes in Jy of these objects (Column 10 of Table 2). 
The bottom left panel is a histogram of the logarithmic distribution of 
maximum 850 $\mu$m intensities in Jy beam$^{-1}$ of these objects (Column 5 
of Table 2).  The bottom right panel is a cumulative logarithmic distribution 
of the alternative 850$\mu$m flux in Jy of these objects.}

\end{figure}

\clearpage

\begin{figure}
\figurenum{12}
\epsscale{1.0}
\vspace*{-0.1cm}
\caption{Histogram of the numbers of Objects in either the Fundamental or 
Extended Map Object Catalogues at a given signal-to-noise level or above.}
\end{figure}

\clearpage

\begin{figure}
\figurenum{13}
\epsscale{1.0}
\vspace*{-0.1cm}
\caption{850 $\mu$m map of the LDN 1551 cloud from the Fundamental Dataset.  
Intensities at 850 $\mu$m range from -0.1 Jy beam$^{-1}$ to 0.42 Jy 
beam$^{-1}$ (greyscale).  Squares denote positions of maximum intensity 
for each object in this map listed in the Fundamental Map Object Catalogue 
(Table 2); these positions correspond to the identifiers of each object 
(Column 1 of Table 2).  The labels denote for each object, the maximum 
signal-to-noise ratio of each object (Column 8 of Table 2).  Dark contours 
delineate the outer boundaries of each object.}
\end{figure}

\clearpage

\begin{figure}
\figurenum{14}
\epsscale{1.0}
\vspace*{-0.1cm}
\caption{850 $\mu$m map of the galaxy M 51 from the Fundamental Dataset.  
Intensities at 850 $\mu$m range from -0.1 Jy beam$^{-1}$ to 0.32 Jy 
beam$^{-1}$ (greyscale).  Symbols, labels and contours are defined as
for Figure 13.}
\end{figure}

\clearpage

\begin{figure}
\figurenum{15}
\epsscale{1.0}
\vspace*{-0.1cm}
\caption{850 $\mu$m map of the NGC 7538 massive star-forming region from 
the Fundamental Dataset.  Intensities at 850 $\mu$m range from -0.1 Jy 
beam$^{-1}$ to 3.2 Jy beam$^{-1}$ (greyscale).  Symbols, labels and 
contours are defined as for Figure 13.}
\end{figure}

\clearpage

\begin{figure}
\figurenum{16}
\epsscale{1.0}
\vspace*{-0.1cm}
\caption{Comparison of maximum 850 $\mu$m intensities in LDN 1551 (crosses) 
and NGC 7538 (triangles) between objects in the Fundamental Dataset maps and 
those in smoothed and rebinned maps by Moriarty-Schieven et al., and Reid \& 
Wilson respectively.  The dashed line denotes the line of equality.}
\end{figure}

\clearpage

\begin{figure}
\figurenum{17}
\epsscale{1.0}
\vspace*{-0.1cm}
\caption{Relative offsets between expected positions of the point-like 
calibrators HL Tau, CRL 618, and CRL 2688 (at center) and the positions of
the pixel of maximum 850 $\mu$m intensity for the corresponding object in
Fundamental Map Object Catalogue (squares).  The dashed circles denote 
angular offsets starting at 1\as\ radius and increasing in steps of 1\as\ 
radius.  The bold dashed circle denotes an angular offset of 6\as, equal 
to 1 $\sigma$ of the narrow component of the {\it unsmoothed} JCMT beam 
at 850 $\mu$m.} 
\end{figure}

\end{document}